\documentclass[
    ,final            
  ]
  {aipproc}

\layoutstyle{6x9}

\begin{document}

\title{Hyperon ordering in neutron star matter}

\author{L. Mornas}{
  address={Departamento de Fisica, Universidad de Oviedo, 
           E-33007 Oviedo (Asturias) Spain}
}

\author{J.P. Su\'arez Curieses}{
  address={Departamento de Fisica, Universidad de Oviedo, 
           E-33007 Oviedo (Asturias) Spain}
}

\author{J. Diaz Alonso{$^*$}}{
  address={LUTH, FRE2462 CNRS, Observatoire de Paris-Meudon, 
            F92195 Meudon , France}             
}

\begin{abstract}
We explore the possible formation of ordered phases in neutron star 
matter. In the framework of a quantum hadrodynamics model where 
neutrons, protons and Lambda hyperons interact via the exchange of 
mesons, we compare the energy of the usually assumed uniform, liquid 
phase, to that of a configuration in  which di-lambda pairs immersed in an 
uniform nucleon fluid are localized on the nodes of a regular lattice. 
The confining potential is calculated self-consistently as resulting 
from the combined action of the nucleon fluid and the other hyperons, 
under the condition of beta equilibrium. We are able to obtain stable 
ordered phases for some reasonable sets of values of the model 
parameters. This could have important consequences on the structure 
and cooling of neutron stars.
\end{abstract}

\maketitle


\section{Introduction}

The equation of state in the interior of neutron stars is usually considered 
to be that of an interacting Fermi liquid in a uniform phase. Under certain 
conditions however, a crystallized phase may be energetically more 
favorable. This in turn can have interesting consequences on the structure 
and evolution of the neutron stars. We may quote as examples thereof 
{\it (i)} triaxial configurations with emission of gravitational waves 
\cite{Haensel:1997},  {\it (ii)} modification of the oscillation modes of
the star \cite{Bonazzola:2002} or {\it (iii)} of the neutrino transport 
properties \cite{BaikoHaensel:1999}.

In a previous paper \cite{PerezGarcia:2002} we presented the results
pertaining to a simplified model where we considered three baryonic 
species: the neutron, proton and Lambda hyperon, interacting with 
each other through the exchange of $\sigma$ and $\omega$ mesons. 
We found that it was possible to find some sets of the model parameters 
such that the energy of a configuration where pairs of $\Lambda$ hyperons 
with antiparallel spins located at the nodes of a cubic lattice, could 
be energetically more favorable than the corresponding liquid phase. 

The model presented in \cite{PerezGarcia:2002} was a strictly minimal one.
Here we would like to perform again these calculations with a more
realistic description of the nuclear interaction.
In order to reproduce the correct values of the incompressibility
modulus and the asymmetry energy, we choose an approach based on the 
Density Dependent Hadron Field  theory (DDRH) developped by the Giessen group 
\cite{Hofmann:2000,Keil:2000,Hofmann:2001}. 
In this model, the nucleon-meson vertices have a functional dependence on 
the density operator. In this way, some basic features of the short-range 
correlations are taken into account. 

\section{Model of baryonic matter}
\subsection{Lagrangian}
We describe the nuclear interaction in the framework of a phenomenological
quantum relativistic model. 
The interaction piece of the Lagrangian density 
in the Density Dependent Hadron Field Theory (DDRH) reads: 
\begin{equation}
{\cal L}_{\rm int}=\sum_{B=N,H} \Gamma_{\sigma B} \overline\psi_B
\, \sigma\, \psi_B - \Gamma_{\omega B} \overline\psi_B
\, \gamma^\mu \omega_\mu\, \psi_B + \Gamma_{\delta B} \overline\psi_B
\, \delta\, \psi_B - \Gamma_{\rho B} \overline\psi_B
\, \gamma^\mu \rho_\mu\, \psi_B
\label{lagrangian}
\end{equation}
where the couplings $\Gamma_{\alpha B}$ are density-dependent through their 
functional  dependence on the baryonic current $J^\mu$.
\begin{equation}
\Gamma_{\alpha B} = \Gamma_{\alpha B}[ \hat J^\mu 
\hat J_\mu] \quad {\rm with} \quad 
\hat J^\mu =\sum_B  \hat{\overline \psi_B} \gamma^\mu \hat \psi_B
\nonumber
\end{equation}
We will consider only one species of hyperons, the $\Lambda$. 
In a realistic model, it can be expected that the $\Sigma$ 
hyperon will also play a role. However, the interaction of this hyperon with 
nuclear matter is less well known, so that it has not yet been implemented 
in the framework of a DDRH-type model.

\subsection{Liquid phase}

In the liquid phase we impose the usual conditions for charge neutrality 
and chemical equilibrium under $\beta$ decay among the baryons (neutron,
proton, $\Lambda$ hyperon) and leptons (electron, muon) taken into
account in our model.
$$ n_e + n_\mu=n_p \, , \qquad
\mu_n=\mu_\Lambda   \, ,\qquad \mu_p-\mu_n=\hat \mu = \mu_e= \mu_\mu
$$
These equations determine the relative fraction of each species of particles.
The equation of state is then obtained from the energy momentum tensor.
For details we refer the reader to \cite{Hofmann:2000,Keil:2000,Hofmann:2001}. 
In particular we will be interested  in comparing the energy density of the 
liquid phase $\rho_L = T^{00}$ to its counterpart in the crystallized phase.

\subsection{Self consistent confining potential}

We now assume that the hyperons are ordered on a lattice and are localized 
in gaussian clouds. We further assume that there are two hyperons with 
antiparallel spins per lattice site . This is partly in order to avoid 
complexities related to spin-spin interaction, but we can also justify 
this choice by arguing that the $\Lambda$-$\Lambda$ interaction, as 
inferred from data on double hypernuclei, is attractive and may even 
favor a bound state ($\sim$ H-dibaryon).  In the spirit of the Sommerfeld
aproximation, we neglect the redistribution of the surrounding nucleons.

The potential energy of an hyperon around a lattice site is
calculated as the sum of the interaction potential energies 
generated by hyperons at other lattice sites $\vec r_i= a (l \vec i +m \vec j
+n \vec k)$.
\begin{equation}
U_{\rm sup}(\vec r) = \sum_i U(\vec r - \vec r_i) \label{confpot}
\end{equation}
In this expression, the potential energy of each hyperon is obtained by taking
the convolution
\begin{equation}
U(\overrightarrow{r})=\frac{2\pi }{r}\int_{0}^{\infty}R\, 
n(R)\, dR \int_{|r-R|}^{|r+R|} V^{OBE} (x)\, x\,  dx
\nonumber
\end{equation}
of a gaussian distribution
\begin{equation}
n(r) = 2 \Psi_{\Lambda}^{*}(r)\Psi_{\Lambda}(r) = 
2 \left( \frac{M_{\Lambda}\nu_{0}}{\pi }\right)^{3/2}e^{- M_{\Lambda}
\nu_{0} r^{2} }
\label{gaussdistr}
\end{equation}
with the elementary one boson exchange potential $V^{OBE} (x)$ 
corrected for the finite size of hyperons by form factors (with cutoffs 
$\Lambda_\sigma$, $\Lambda_\omega$) and for relativistic effects in the 
momentum expansion. Note that in the present configuration, only the 
central and spin-spin components contribute.

The potential calculated in this way is then approximated by a parabola 
$U_{\rm par}$ parameterized by its depth $U_0$ and the frequency of 
oscillation of the  hyperons in the potential well $\nu_0=\sqrt{ \nabla^{2} 
U_{\rm par}(\overrightarrow r )|_{\overrightarrow r =0} / M_\Lambda}$.
The width $\Gamma= 1/\sqrt{M_\Lambda\, \nu_0}$ of the gaussian distribution
\eqref{gaussdistr} is determined self-consistently by the frequency of 
this harmonic oscillator.

\subsection{Equation of state of the ordered phase}

We again impose the conditions for $\beta$ equilibrium
\begin{equation}
\mu_n=\mu_p+\mu_e\, , \quad  \mu_n=\mu_\Lambda\, , \quad \mu_e=\mu_\mu
\label{eqbeta}
\end{equation}
with the chemical potentials given by
\begin{eqnarray}
\mu_{p} &=&\sqrt{p_{fp}^{2}+M_p^{2}}
+\Gamma_{\omega N}( <\omega^{0}> + <\omega_{\Lambda}>_s) 
+\Gamma_{\rho N} <\rho^{0}> +\Sigma^N_{\rm rearr} \nonumber \\
\mu_{n} &=&\sqrt{p_{fn}^{2}+M_n^{2}}
+\Gamma_{\omega N}( <\omega^{0}> + <\omega_{\Lambda}>_s) 
-\Gamma_{\rho N} <\rho^{0}> +\Sigma^N_{\rm rearr}\nonumber \\
\mu_{\Lambda} &=&  M_{\Lambda}+U_0+\frac{5}{2} \nu_{0}
+\Gamma_{\omega \Lambda} <\omega^{0}> +\Sigma^\Lambda_{\rm rearr}
\label{chempot}
\end{eqnarray}
The nucleons evolve in constant mean fields consisting of two contributions: 
To the field produced by the homogeneous nucleon fluid ({\it e.g.} 
$\omega^{0}$), we add the spatial average of the fields generated by the 
periodic hyperon distribution  ({\it e.g.} $<\omega_{\Lambda}>_s$). 
The chemical potential of the hyperons is now determined by the parameters 
of the confining potential. The rearrangement terms $\Sigma_{\rm rearr}$ 
contain derivatives of the couplings $\Gamma_{\alpha  B}$ with respect 
to the density.

We further impose the conditions of charge neutrality $n_\mu + n_e = n_p $, 
and express the baryonic density $n_B=n_n+n_p+n_\Lambda$ as the sum of the 
nucleon densities and the hyperon density, the latter being related to the lattice 
parameter $a$ by $n_\Lambda= 2 /a^3$.
Finally, we write the defining equations for the effective baryon masses and
the self consistent equations obeyed by the fields 
$\sigma$, $\omega^0$, $\rho^0$ and $\delta$.
\begin{eqnarray}
M_p &=& m_N - \Gamma_{\sigma N}( <\sigma> +  
<\sigma_\Lambda>_{s}) - \Gamma_\delta  <\delta> \nonumber \\ 
M_n &=& m_N - \Gamma_{\sigma N} ( <\sigma> +  
<\sigma_\Lambda>_{s}) + \Gamma_\delta  <\delta> \nonumber \\
M_\Lambda &=& m_\Lambda-\Gamma_{\sigma \Lambda} <\sigma> \nonumber \\
& &  < \sigma> = \Gamma_{\sigma N} (n^{(s)}_p + n^{(s)}_n) 
= {\Gamma_{\sigma N} \over m_{\sigma}^2 } 
\sum_{i=p,n} { M_i \over 2\pi^2} \left[  p_{fi} 
\varepsilon_{fi}  - M_{i}^{2} \ln \left( { p_{fi}+
 \varepsilon_{fi} \over  M_{i}} \right) 
\right]   \nonumber \\
&& < \delta> = \Gamma_{\delta N} (n^{(s)}_p - n^{(s)}_n) \ , \qquad 
<\sigma_{\Lambda}>_{s} = {\Gamma_{\sigma \Lambda} \over 
m_{\sigma }^{2}} n_\Lambda  \nonumber \\
&& <\omega^{0}>= {\Gamma_{\omega N} \over m_{\omega }^{2}} (n_n+n_p)= 
 {\Gamma_{\omega N} \over m_{\omega }^{2}}\left( 
\frac{ p_{fp}^{3}}{3\pi ^{2}} + \frac{{ p_{fn}}^{3}}
{3\pi ^{2}}\right)    \nonumber \\
&& <\rho^{0}>= {\Gamma_{\rho N} \over m_\rho^2} 
(n_p-n_n)\ , \qquad  <\omega_{\Lambda}>_{s} = {\Gamma_{\omega \Lambda} \over 
m_{\omega }^{2}} n_\Lambda  
\label{effmasses}
\end{eqnarray}
\vskip -0.15cm 
The equations for chemical equilibrium 
\eqref{eqbeta},\eqref{chempot},\eqref{effmasses} 
again determine the chemical composition of the matter at a
given baryonic density. They have to be solved self consistently with
the equations for the confining potential \eqref{confpot}, to which
they are interrelated through the effective hyperon mass, $M_\Lambda$
the lattice parameter $a$, the depth $U_0$ and the oscillation frequency 
$\nu_0$.

\section{Numerical results}

We will use the rational parametrization of the coupling constants
suggested in Ref. \cite{TypelWolter:1999} with the parameters of Ref. 
\cite{Hofmann:2000}
\vskip -0.25cm
\begin{equation}
\Gamma_{\alpha N} (n_B) = 
      a_\alpha \left[ { 1 + b_\alpha \left( {n_B \over n_{\rm sat}} +
      d_\alpha \right) \over 1 + c_\alpha \left( {n_B \over n_{\rm sat}} +
      e_\alpha \right)} \right] \ , \qquad \Gamma_{\alpha \Lambda} ( n_B) =
      x_{\alpha \Lambda} \Gamma_{\alpha N} (n_B) 
\label{couplconst}
\end{equation}
\vskip -0.1cm\noindent
This function is chosen so as reproduce parameter free Dirac-Brueckner 
calculations in the nucleon sector. The parameters $x_{\alpha B}$ are 
adjusted in order to reproduce the binding energy of the $\Lambda$
in hypernuclei. 
The coupling constants used in the present work correspond to the ``model 1
with DD phenomenological'' parametrization of Reference \cite{Hofmann:2001}.
We would like to stress the fact that, in contrast to the calculation
performed in \cite{PerezGarcia:2002}, we have essentially no free parameters
in the results presented in this work.

When numerical convergence is reached, we obtain the parameters
$U_0$, $ \nu_0$, $ M_i$, $p_{Fi}$, {\it etc.} which fully determine the 
thermodynamical state of the system. We now are in a position to calculate
the energy density of the crystallized phase. It is obtained as
\vskip -0.2cm
\begin{equation}
\rho_C = \rho_p + \rho_n + \rho_e + \rho_\mu + \rho_\Lambda 
+ \rho_{\rm fields} 
\end{equation}
\vskip -0.2cm
\begin{eqnarray}
\rho_\Lambda &=& (M_{\Lambda} + \frac{3}{2}\nu_{0} + \Gamma_{\omega \Lambda}
<\! \omega^{0} \! > + U_{0})\, n_{\Lambda} \nonumber \\
\rho_{\rm fields}\!\!\!\! &=&\!\!\!\! \frac{\Gamma_{\sigma N}}{2}\! 
              \left( \! <\!\!\sigma\!\!> \!+\! 
             <\!\! \sigma_\Lambda \!\! >_s \! \right) (n^{(s)}_p + n^{(s)}_n) 
             \!+\! \frac{\Gamma_{\sigma \Lambda}}{2}\!  <\!\!\sigma\!\!>  
             < \!\! n^{(s)}_{\Lambda} \!\! >_s
            + \frac{\Gamma_{\delta N}}{2}\! <\!\!\delta\!\!>\!   
                (n^{(s)}_p - n^{(s)}_n)   \nonumber \\ 
         & + &\!\!\!\!\frac{\Gamma_{\omega N}}{2} \left(  <\!\!\omega^0\!\!> 
               \!+\! <\!\! \omega_\Lambda \! >_s \right) (n_n + n_p) 
               \!+\! \frac{\Gamma_{\omega \Lambda}}{2} <\!\!\omega^0\!\!> 
               <\!\! n_\Lambda \!\! >_s 
               \!+\! \frac{\Gamma_{\rho N}}{2}  <\!\!\rho^0\!\!>\! (n_p - n_n)
\nonumber  
\end{eqnarray}
This energy is then compared to the energy density of the corresponding
liquid phase $\rho_L$. The result is displayed in Figure 1.
With the chosen model parameters we obtained an ordered phase which
is energetically more favorable than the liquid one above the
threshold for hyperon production $n_{\rm th}=1.95\, n_{\rm sat}$.
The shape of this curve is similar to the one which was obtained for 
parameter set C in \cite{PerezGarcia:2002}, with however a somewhat 
smaller energy gain. The stability of the solid phase increases at high 
density.

\begin{figure}
\begin{minipage}{12.8cm}
\null\hskip -1.2cm
\mbox{%
\parbox{6.2cm}{
  \includegraphics[height=.25\textheight]{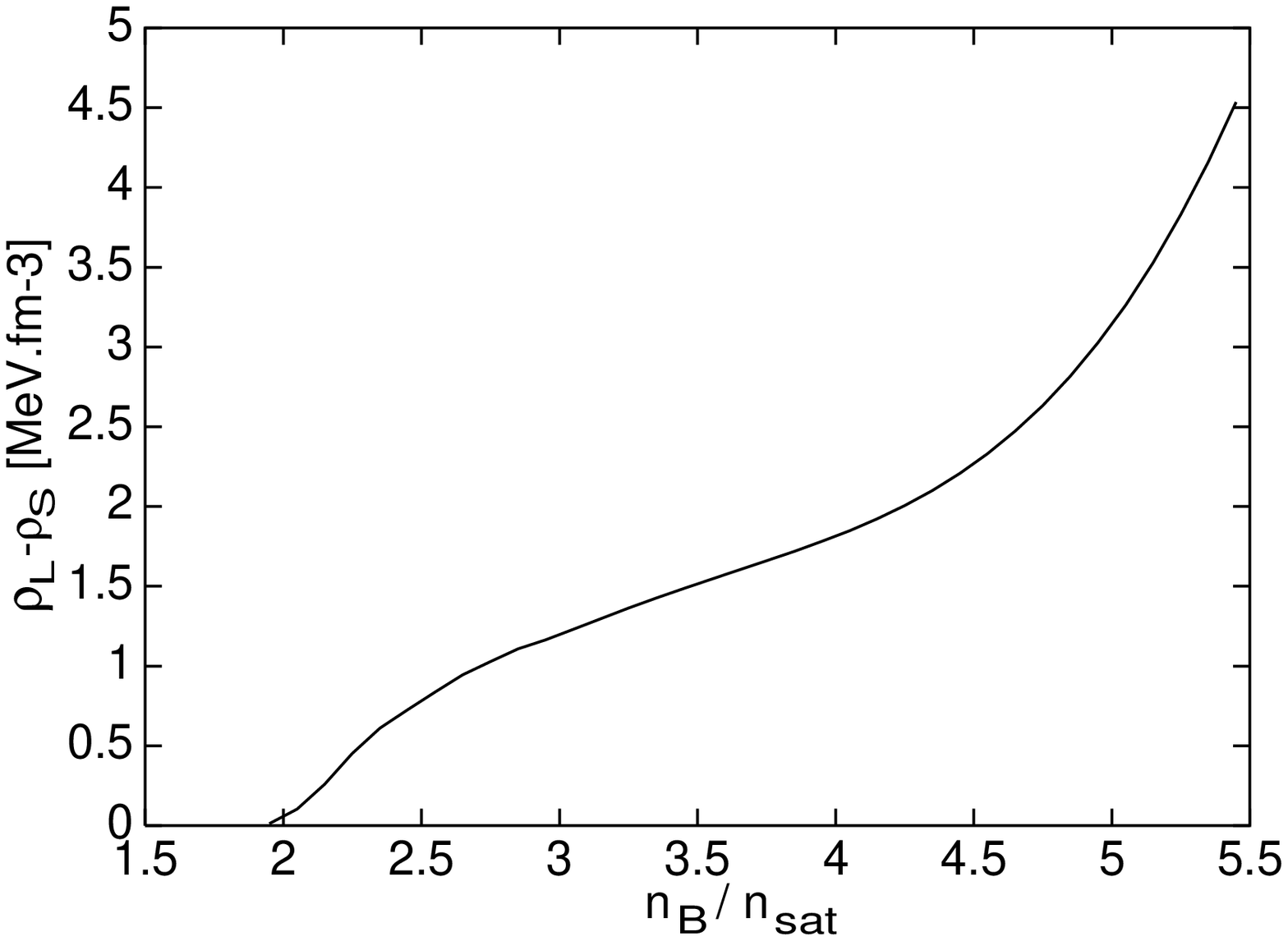}}
\parbox{0.4cm}{\phantom{space}}
\parbox{6.2cm}{
  \includegraphics[height=.25\textheight]{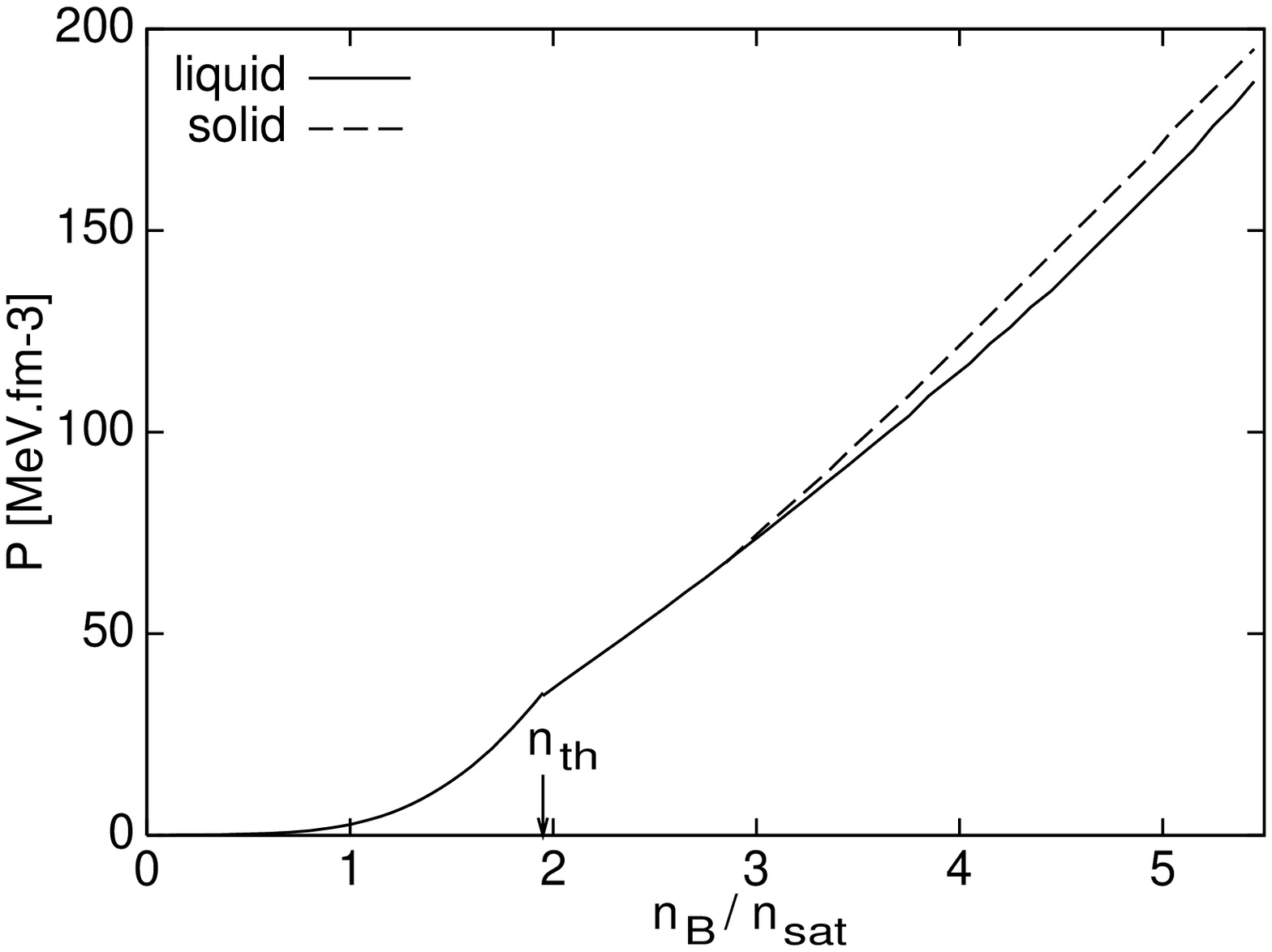}}
}
\vskip 0.1cm
\null\hskip -0.5cm
\mbox{%
\parbox{6.3cm}{\small{{\bf Figure 1.}
   Energy difference between the liquid and crystallized phase,
   as a function of baryonic density}} 
\parbox{0.2cm}{\phantom{space}}
\parbox{6.3cm}{\small{{\bf Figure 2.}
   Pressure in the liquid and the crystallized phases,
   as a function of baryonic density.}}
} 
\end{minipage}
\end{figure}

Finally, we obtain the pressure from the thermodynamical relation
\begin{equation}
P=n_{B}^{2}{\frac{\partial (\rho /n_{B})}{\partial n_{B}}}
\end{equation}
The result is displayed in Figure 2. 
In contrast with the results obtained  for parameter set C in 
\cite{PerezGarcia:2002}, the equation of state of the ordered phase
is now slighlty harder than the liquid one.

\section{Conclusion}

We investigated the possible formation of an ordered phase of the baryonic
matter present in the core of the neutron stars, in which the hyperons
are located on the nodes of a cubic lattice. The model presented in this 
short contribution is a first upgrade of our original barebones model
\cite{PerezGarcia:2002}. We now have a more realistic description of the 
nuclear interaction, in particular of the nucleon sector. Our main 
conclusion is that the findings of \cite{PerezGarcia:2002} are confirmed, 
namely the ordered phase can be energetically more favorable for some 
choices of the model parameters. As could be expected, our result is found 
to depend more strongly on the features of the  -- largely unknown -- 
in-medium hyperon-hyperon potential than on the properties of the nucleon 
background such as incompressibility modulus and asymmetry. A more
systematic study with several alternative parametrizations is 
underway and will be presented in a further publication.

The model which we have presented here can be improved in several ways.
In order to be consistent with the DDRH picture, we described the 
hyperon-hyperon interaction by sigma and omega exchange with density
dependent couplings in the form $\Gamma_{\alpha \Lambda} (n_N,n_\Lambda)
= x_{\alpha \Lambda} \Gamma_{\alpha N} (n_N +n_\Lambda)$ with $\alpha 
\in \{ \sigma, \omega \}$. On the other hand, some authors 
\cite{Schaffner:1994} also introduce additional $\sigma*$ 
and $\phi$  exchange in order to reproduce the attractive $\Lambda-\Lambda$ 
interaction inferred from the available data on double hypernuclei.
Furthermore, parametrizations of the free hyperon-nucleon and
hyperon-hyperon potentials consider the exchange of {\it e.g.} kaons or 
$\eta$ which are also not present in our calculations.

Ideally one could extract an in-medium hyperon-hyperon potential 
from  Brueckner-type calculations and test the result comparing with  
the available hypernuclei data. Whereas several non relativistic 
calculations exist in the litterature \cite{Baldo:2000,Vidana:2000}, 
the corresponding calculations have not been performed in a relativistic 
framework. In particular a full fledged relativistic calculation
including all hyperons (not only $\Lambda$ but also $\Sigma$ and $\Xi$)
and a subsequent DDRH-type parametrization would be especially welcome.
It would allow us to have a greater variety of lattice configurations 
involving in particular the $\Sigma^-$ hyperon. Indeed the $\Sigma^-$ 
is formed in the liquid phase at the same density as the $\Lambda$ 
in most neutron star matter calculations.

An other direction of research would be to release the Sommerfeld 
approximation and take into account the redistribution of nucleons.
This involves considering screening correlations of the RPA type
to the potential.

{\bf Acknowledgment:} This work was partially supported by project 
MCT-00-BFM-0357.




\bibliographystyle{aipproc}   

\bibliography{hyperons}

\IfFileExists{\jobname.bbl}{}
 {\typeout{}
  \typeout{******************************************}
  \typeout{** Please run "bibtex \jobname" to optain}
  \typeout{** the bibliography and then re-run LaTeX}
  \typeout{** twice to fix the references!}
  \typeout{******************************************}
  \typeout{}
 }

\end{document}